\documentclass[aps,pra,showpacs,twocolumn,amsmath,amssymb,footinbib]{revtex4}

\usepackage[english]{babel}
\usepackage{latexsym}
\usepackage{graphics,psfrag}
\usepackage{epsfig}
\usepackage[T1]{fontenc}
\usepackage[usenames,dvipsnames]{color}
\usepackage{ulem}

\def\be{\begin{equation}}
\def\ee{\end{equation}}
\def\bea{\begin{eqnarray}}
\def\eea{\end{eqnarray}}
\def\bi{\begin{itemize}}
\def\ei{\end{itemize}}
\def\bin{\begin{enumerate}}
\def\ein{\end{enumerate}}
\def\la{\langle}
\def\ra{\rangle}

\newcommand{\vect}[1]{\mathbf{#1}}

\begin{document}

\title{Frustration and time reversal symmetry breaking for Fermi and Bose-Fermi systems}

%%%%%%%%%%%%%%%%%%%%%%%%%%%%%%%%%%%%%%%%%%%%%%%%%%%%%%%%%%%%%%%%%%%%%%%%%%%%%%%
\author{Krzysztof Sacha  }
\affiliation{
Instytut Fizyki imienia Mariana Smoluchowskiego, 
Uniwersytet Jagiello\'nski, ulica Reymonta 4, PL-30-059 Krak\'ow, Poland}

\affiliation{
Mark Kac Complex Systems Research Center,  
Uniwersytet Jagiello\'nski, Krak\'ow, Poland}

\author{Katarzyna Targo\'nska}

\affiliation{
Instytut Fizyki imienia Mariana Smoluchowskiego, 
Uniwersytet Jagiello\'nski, ulica Reymonta 4, PL-30-059 Krak\'ow, Poland}
\affiliation{
Mark Kac Complex Systems Research Center, 
Uniwersytet Jagiello\'nski, Krak\'ow, Poland}

\author{Jakub Zakrzewski}
\affiliation{
Instytut Fizyki imienia Mariana Smoluchowskiego, 
Uniwersytet Jagiello\'nski, ulica Reymonta 4, PL-30-059 Krak\'ow, Poland}

\affiliation{
Mark Kac Complex Systems Research Center, 
Uniwersytet Jagiello\'nski, Krak\'ow, Poland}

\date{\today}

\begin{abstract}
The modulation of an optical lattice potential that breaks time-reversal symmetry enables the realization  of complex tunneling amplitudes in the corresponding tight-binding model. For a superfluid Fermi gas in a triangular lattice potential with complex tunnelings the pairing function acquires a complex phase, so the frustrated magnetism of fermions can be realized.
Bose-Fermi mixture of bosonic molecules and unbound fermions in the lattice shows also an interesting behavior. Due to boson-fermion coupling, the fermions become slaved by the bosons and the corresponding pairing function takes the complex phase determined by bosons. In the presence of bosons the Fermi system can reveal both gap and gapless superfluidity.
\end{abstract}


\maketitle
%%%%%%%%%%%%%%%%%%%%%%%%%%%%%%%%%%%%%%%%%%%%%%%%%%%%%%%%%%%%%%%%%%%%%%%%%%%%%%%
\section{Introduction}

 Cold atoms in optical lattices provide a unique medium for mimicking effects known from other areas of physics. This is primarily due to a great flexibility and a precise manipulation of the cold atomic system \cite{jaksch05,lewen07,bloch08}. Atoms of  a fermionic or bosonic character may be placed in an optical lattice potential whose geometry may be easily controlled by changing directions and/or polarizations of laser beams. Interactions between atoms may be controlled via magnetic, optical or microwave Feshbach resonances \cite{chin10,hanna10}. 
The change in  the depth of the optical lattice modifies primarily the tunneling between lattice sites
(having also effect on the effective interaction strength) enabling e.g.
 the superfluid-Mott insulator quantum phase transition in the optical realization of the Bose-Hubbard model as proposed by Jaksch and Zoller \cite{jaksch98} and subsequently demonstrated in Ref.~\cite{greiner02}.

Another spectacular way of controlling the tunneling has been proposed by Eckardt, Weiss and Holthaus \cite{eckardt05}. Fast periodic modulations of the optical lattice allow for an effective, time-averaged tunneling to be totally suppressed keeping the depth of the lattice potential unchanged. By varying the strength of the modulation one can induce the superfluid-Mott insulator quantum phase transition \cite{eckardt05} as verified experimentally a few years later \cite{lignier07,zenesini09}.
  Importantly, not only the magnitude, but
also the sign of the tunneling amplitudes can be altered  using this approach.
This concept has been utilized  in a recent proposition to create frustrated magnetism with cold bosons in a triangular lattice \cite{eckardt10}
later implemented in fascinating experiments of the Hamburg group \cite{sengstock11}.
 
The effective tunneling caused by periodic lattice modulations is adequately
 explained in the framework of the Floquet theory \cite{eckardt05} for  the periodically time-dependent Hamiltonians. 
The properties of the corresponding quasi-energies spectra are known to depend on the global symmetries of the Hamiltonian \cite{sacha99,sacha00,sacha01}. 
Employing the similar idea we demonstrate that periodic perturbations that break the time-reversal invariance (TRI) can change not only the sign of the tunneling amplitudes, but may also induce their complex values.

In the following, we concentrate on superfluid fermions in the Bardeen-Cooper-Schriffer (BCS) regime with broken TRI. We show
 that typically a pairing function for $s$-wave interactions acquires a complex phase which may be controlled by the TRI breaking mechanism considered in the present paper (for a discussion of $p$-wave orbitals see
 \cite{brayali09}). We also point out that in the Bose-Fermi mixture with broken TRI the complex ground state of bosons  affects the Fermi pairing function. 
The effect is reminiscent of the disorder induced phase control in such mixtures as discussed by one of us recently \cite{niederberger09}. 
In our case the phase control is not due to a disorder, but due to a control over the tunneling mechanism and the TRI breaking.  

\section{Breaking the time-reversal symmetry}

\subsection{One-dimensional optical lattice}

Let us begin, for simplicity, with a single particle in an one-dimensional (1D) optical lattice potential driven by a double harmonic perturbation. The Hamiltonian of the system reads
\be
H_0=\frac{p^2}{2m}+V(x)+K_1x\cos(\omega t)+K_2x\cos(2\omega t+\varphi),
\label{1p}
\ee
where $V(x)=V(x+a)$ is an optical lattice potential with the lattice constant $a$, $K_{1,2}$ stand for strengths of the driving at a basic frequency $\omega$ and its second harmonic, respectively.
The Hamiltonian is time-periodic, i.e. $H_0(t+2\pi/\omega)=H_0(t)$, and the Floquet theorem
\cite{floquet883,shirley65,zeldovich66} guarantees that 
so called the Floquet Hamiltonian:
\be
{\cal H}=H_0-i\hbar\partial_t
\ee
is diagonalized by periodic functions.
Eigenvalues of the $\cal H$ are referred to quasi-energies of the system analogous to quasi-momenta in a solid state physics. They are defined modulo $\hbar\omega$ and it is sufficient to consider a single Floquet zone (an analog of the Brillouin zone). Periodic eigenfunctions of the Floquet Hamiltonian can be expanded in a Fourier series, i.e. in a basis where time $t$ can be considered as an additional {\it degree of freedom}. Let us define basis vectors
\bea
\phi_{j,m}(x,t)&=&\exp\left\{-ix\left[\frac{K_1}{\omega}\sin(\omega t)+\frac{K_2}{2\omega}\sin(2\omega t+\varphi)\right]\right\}
\cr && \cr && \times \exp\left(im\omega t\right)W_j(x),
\label{basis}
\eea
which fulfill 
\bea
\la\la\phi_{j',m'}|\phi_{j,m}\ra\ra&=&\frac{\omega}{2\pi}\int\limits_0^{2\pi/\omega}dt\int dx \;
\phi_{j',m'}^*\phi_{j,m}
\cr &&
\cr &=&\delta_{j',j}\;\delta_{m',m},
\eea
where $m$ denotes a Fourier component and $W_j(x)=W(x-x_j)$ is a Wannier function of the lowest energy band localized on the $j$-th lattice site. The first phase factor on the r.h.s. of Eq.~(\ref{basis}) corresponds to the unitary transformation which allows us to switch from the length gauge to the velocity gauge using  quantum optics language. The matrix of the Floquet Hamiltonian consists of diagonal $m$-blocks in the basis (\ref{basis}). The blocks are very weakly coupled among themselves provided the driving frequency $\omega$ is very high. Using the tight-binding approximation and taking into an account nearest neighbor tunneling only the diagonal blocks become  
\bea
\la\la \phi_{j',m}|{\cal H}|\phi_{j,m}\ra\ra&=&-J_{\rm eff}\;\delta_{j',j+1}-J^*_{\rm eff}\;\delta_{j',j-1} \cr && \cr &&
+(m\omega+E_0)\;\delta_{j',j}, 
\eea
where $E_0=\la W_j|(p^2/2m+V)|W_j\ra$
%=\int dxW^*(x-x_{j})\left[\frac{p^2}{2m}+V\right]W(x-x_j)$ 
and the effective tunneling amplitude
\bea
J_{\rm eff}=J\displaystyle\sum\limits_{k=-\infty}^{\infty}{\cal J}_{2k}(s_1){\cal J}_{k}(s_2)e^{ik\varphi},
\label{jeff}
\eea
where 
\be
s_i=\frac{aK_i}{\omega},
\label{si}
\ee
are dimensionless strengths of the first ($i=1$) and the second ($i=2$)
harmonic,
the bare tunneling amplitude is $J=-\la W_{j+1}|(p^2/2m+V)|W_j\ra$ and ${\cal J}_n$ is the ordinary Bessel function. If $\hbar\omega\gg J$ the description of a single particle system may be restricted to a single diagonal block
\be
H_{\rm eff}=\la\la \phi_{j',0}|{\cal H}|\phi_{j,0}\ra\ra=-J_{\rm eff}\;\delta_{j',j+1}-J^*_{\rm eff}\;\delta_{j',j-1},
\label{heff}
\ee
where the constant term $E_0$ has been omitted.

\begin{figure}%[h]
\centering
\includegraphics*[scale=0.33]{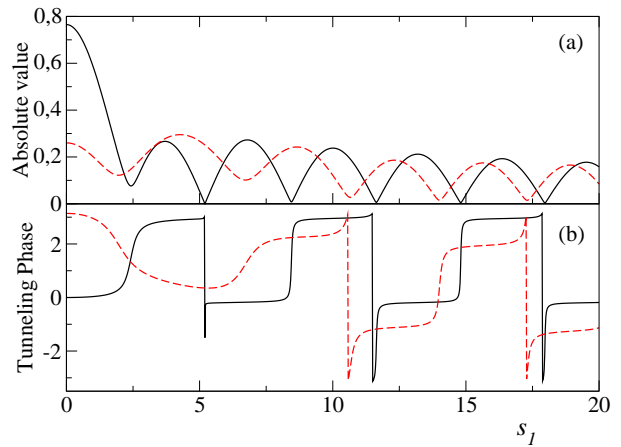}
\begin{center}
\end{center}
\caption{(Color online) 
The absolute value (top), $A$, and the complex phase $\varphi_J$ (bottom) of the effective tunneling amplitude, $J_{\rm eff}/J=A\exp(i\varphi_J)$, Eq.~(\ref{jeff}), for a double harmonic modulation of the optical lattice potential as a function of the dimensionless strength $s_1$ of $\omega$ component
for $s_2=1$ [see (\ref{si})], $\varphi=0.2$ (black solid lines) and  $s_2=3$, $\varphi=0.5$ (red, dashed lines).}
\label{effrate}
\end{figure}

If there is only one harmonic present in (\ref{1p}) or the phase $\varphi=0$,  the Floquet Hamiltonian is time-reversal invariant and $\cal H$ is represented by a real symmetric matrix in a generic basis \cite{haake}. Then, the effective tunneling amplitude (\ref{jeff}) is real. Single harmonic driving has been used to change the interaction from ferromagnetic (positive $J_{\rm eff}$) to antiferromagnetic (negative $J_{\rm eff}$) and to realize frustrated magnetic phases \cite{eckardt10,sengstock11}. By breaking TRI we are able to realize nearly arbitrary complex values of the tunneling amplitude $J_{\rm eff}=|J_{\rm eff}|e^{i\varphi_J}$. In Fig.~\ref{effrate} we present the absolute value $|J_{\rm eff}|$ and the phase $\varphi_J$ as a function of the parameter $s_1$. 

The eigenstates of the Hamiltonian (\ref{heff}) are the Bloch waves $\psi_j=e^{ikx_j}/\sqrt{N_s}$, where $N_s$ is the number of lattice sites, with the dispersion relation 
$E(k)=-2|J_{\rm eff}|\cos(ka-\varphi_J)$.
Single harmonic driving allows for $\varphi_J=0$ or $\pi$ and thus for the ground state with $k=0$ or with $k$ at the edge of the first Brillouin zone. The ground state of the system with broken TRI may correspond to any value of $k$.

We have concentrated on a 1D problem. However, a similar control of phases of the tunneling amplitudes can be also realized in higher dimensions. Indeed, the modulations applied to the lattice along orthogonal axes enable us to produce arbitrary tunnelings along the corresponding directions  \cite{eckardt10}. In the following we will focus on a 2D triangular lattice.

%%%%%%%%%%%%%%%%%%%%%
%%%%%%%%%%%%%%%%%%%%%

\subsection{Two-dimensional triangular optical lattice}
  
\begin{figure}%[h]
\centering
\includegraphics*[scale=0.25]{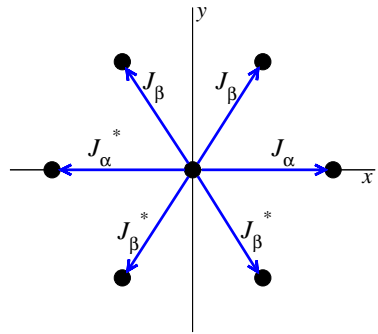}
\caption{(Color online) 
Triangular Bravais lattice points (black circles) and amplitudes $J_{\alpha,\beta}$ corresponding to tunneling from a lattice point to the nearest neighbors.
}
\label{trlatt}
\end{figure}

\begin{figure}%[h]
\centering
\includegraphics*[scale=0.9,angle=-0]{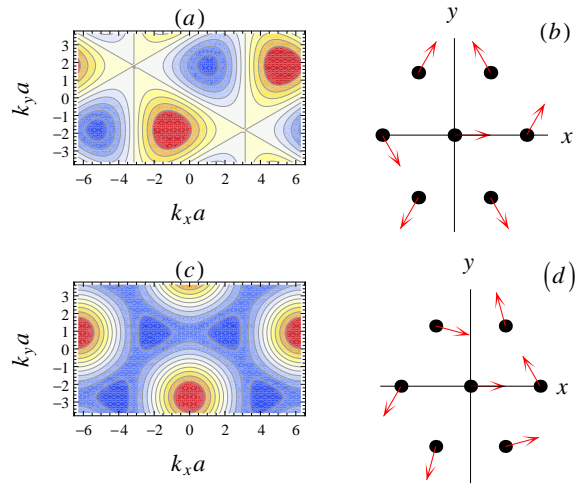}
\caption{(Color online) 
Contour plots of the dispersion relation Eq.~(\ref{ek}) for $|J_{\alpha}|=|J_{\beta}|$
and $\varphi_\alpha=\varphi_\beta=\pi/2$ (a) and $\varphi_\alpha=\pi$ and $\varphi_\beta=\pi/4$ (c); cool colors indicate regions around energy minima. In right panels directions of arrows  indicate phases $e^{i\vect k\cdot \vect r_i}$  where $\vect k$ corresponds to a minimum of the dispersion relation. 
Specifically $\vect ka=\left(\frac{\pi}{3},\frac{\pi}{\sqrt{3}}\right)$  for the minimum in (a) and
 $\vect ka=\left(+\frac{2\pi}{3},\frac{\pi}{2\sqrt{3}}\right)$ for one of the two degenerate, non-equivalent minima in (c). Arrows in panel (b) and (d)  are related to (a) and (c), respectively. 
}
\label{disrel}
\end{figure}

The triangular optical lattice can be realized experimentally by means of three laser beams. Single harmonic modulations of the lattice along the two orthogonal directions allows one to control the sign of the tunneling amplitudes of particles loaded in the lattice. This setup has been used in the Hamburg experiments that demonstrated the frustrated classical magnetism \cite{sengstock11}. With the help of the double harmonic modulations we are able to realize any phase of the tunneling amplitudes
\bea
J_\alpha=|J_\alpha|e^{i\varphi_\alpha}, \quad 
J_\beta=|J_\beta|e^{i\varphi_\beta},
\label{rates}
\eea 
 see Fig.~\ref{trlatt}.
Eigenstates of a single particle in such a lattice are Bloch waves with the dispersion relation
\bea
E(\vect k)&=&-2|J_\alpha|\cos(k_xa-\varphi_\alpha)
\cr &&
-2|J_\beta|\left\{
\cos\left[\left(\sqrt{3}k_y+k_x\right)\frac{a}{2}-\varphi_\beta\right]\right.
\cr &&
\left.+\cos\left[\left(\sqrt{3}k_y-k_x\right)\frac{a}{2}-\varphi_\beta\right]\right\}.
\label{ek}
\eea
We induce a shift of the dispersion relation along the $k_y$ direction in the reciprocal space by changing the value of $\varphi_\beta$ (with the other parameters fixed). The modification of $\varphi_\alpha$ alters the structure of the dispersion relation. It can reveal a doubly degenerate ground state for $\varphi_\alpha=\pi$. The presence of such a degeneracy has been observed experimentally in a Bose system \cite{sengstock11}. For example, for $J_\alpha=J_\beta=-|J_\beta|$ the system in most experimental realizations chooses spontaneously one of the two ground states. With the double harmonic modulation breaking a TRI, the two degenerate minima for $\varphi_\alpha=\pi$ can be moved arbitrarily along the $k_y$ direction with a change of $\varphi_\beta$, see Fig.~\ref{disrel}.

In the Hamburg experiment a Bose-Einstein condensate (BEC) has been prepared in a triangular lattice. Although, in that case particle interactions are present,  the ground state is still determined by the single particle dispersion relation (\ref{ek}). Indeed, assuming an homogeneous system (which is a good approximation of the experimental situation) the solution of the Gross-Pitaevskii equation has the  chemical potential given by $\mu_B=E(\vect k)+n_BU_B$, where 
$U_B$ characterizes the on-site particle interactions and $n_B$ is an average number of bosons per a lattice site. 
We would like to stress, that in the presence of the interactions, the restriction to a single block of the Floquet Hamiltonian like in (\ref{heff}) is valid provided $\hbar\omega\gg U_B$ \cite{eckardt05}. On the other hand, $\hbar\omega$ must be much smaller than the energy separation between bands of the periodic lattice problem for the description limited to the lowest band to be valid.

%%%%%%%%%%%%%%%%%%%%%
%%%%%%%%%%%%%%%%%%%%%

%%%%%%%%%%%%%%%%%%%%%
%%%%%%%%%%%%%%%%%%%%%

%%%%%%%%%%%%%%%%%%%%%
\begin{figure}%[h]
\centering
\includegraphics*[scale=0.6,angle=-0]{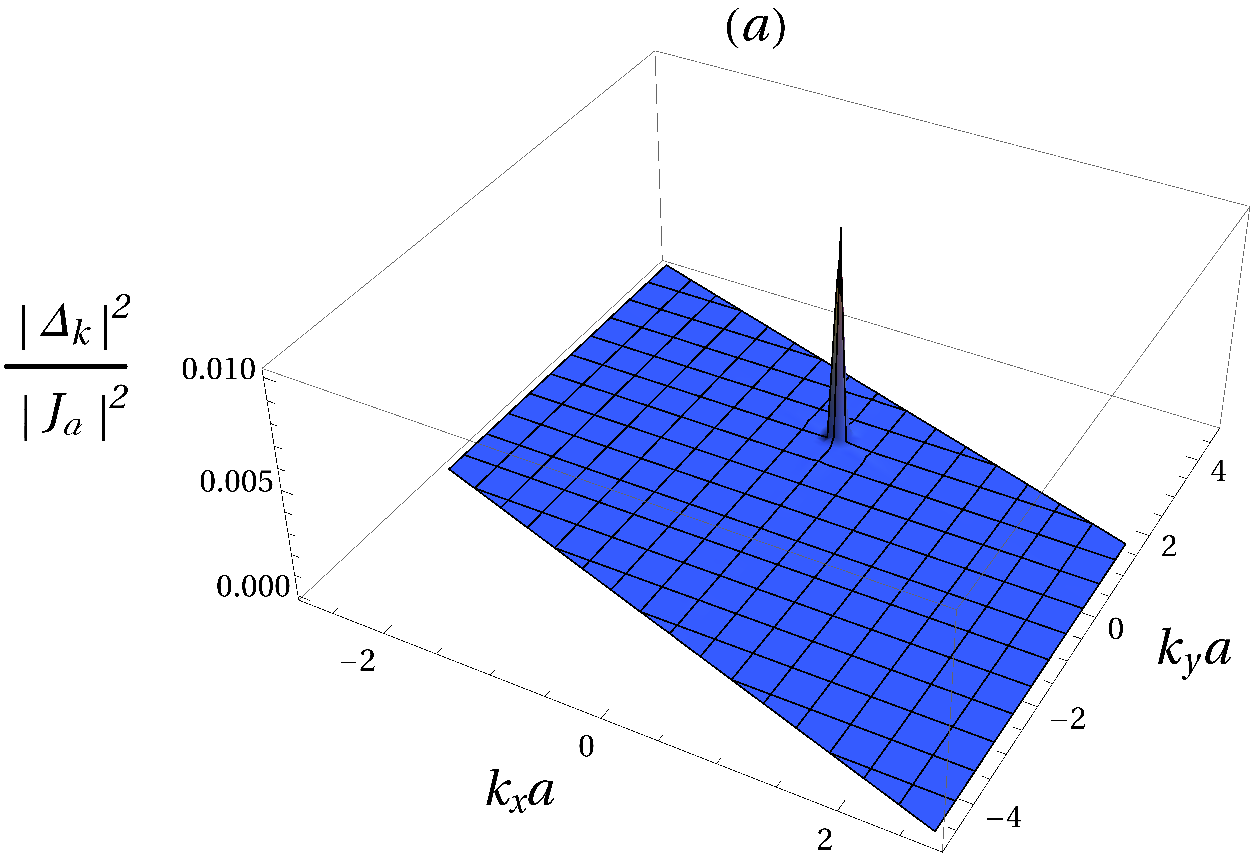}
\includegraphics*[scale=0.6,angle=-0]{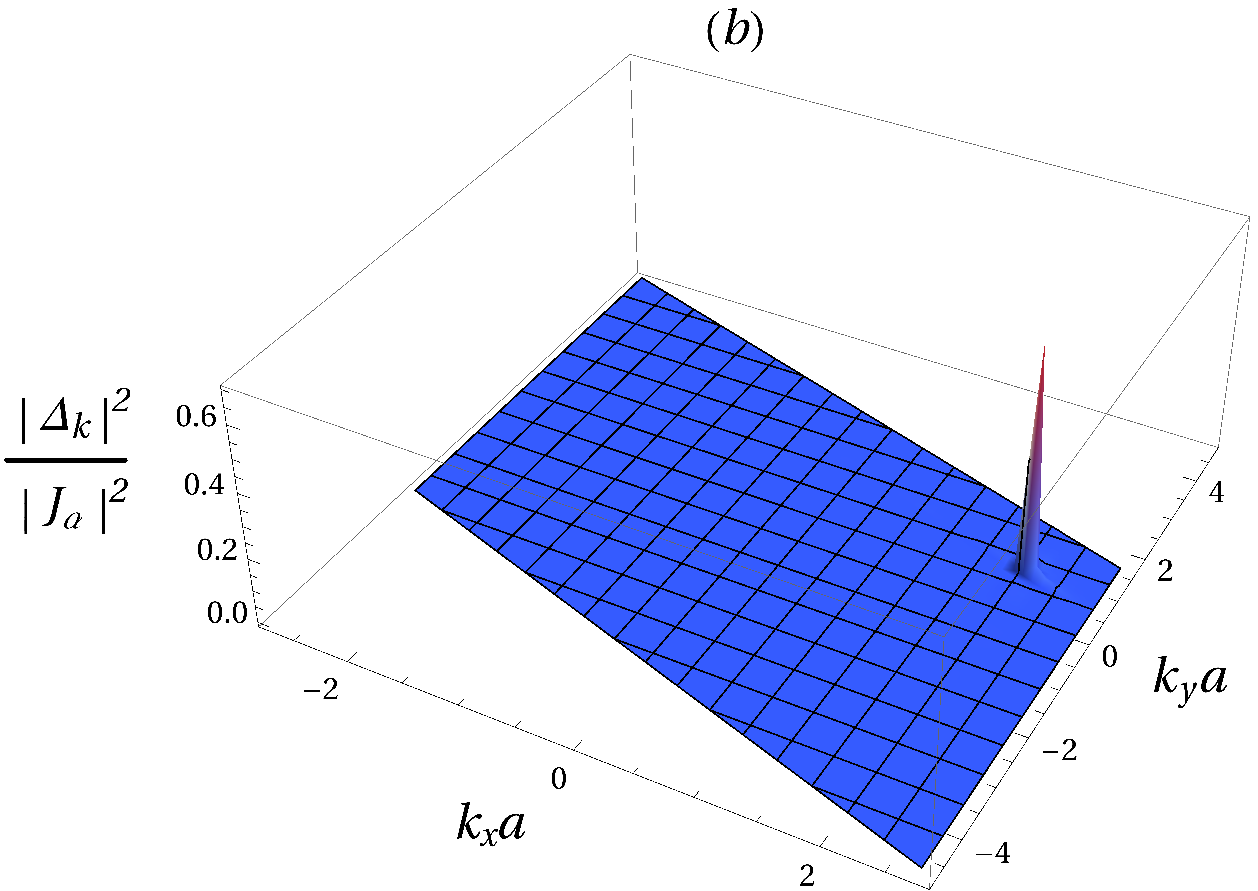}
\caption{(Color online) 
Modulus squared of the Fourier transform of the BCS pairing function $|\Delta_{\vect k}|^2$, where $\Delta_{\vect k}=\sum_{i}\Delta_ie^{-i\vect k\cdot\vect r_i}/\sqrt{N_s}$, obtained numerically for a finite system of $60\times60$ lattice sites for $|J_\alpha|=|J_\beta|$, $\varphi_\alpha=\pi$, $\varphi_\beta=\pi/4$, $U/|J_\alpha|=2$ and $\mu=0$. Panel (a) shows $|\Delta_{\vect k}|^2$ corresponding to the ground state of the isolated Fermi system. Panel (b) presents similar results but for the Fermi system coupled to the Bose-Einstein condensate, with wavefunction $\psi_i=\sqrt{n_B}e^{i\vect q_0\cdot\vect r_i}$, where $\gamma\sqrt{n_B}/|J_\alpha|=2.3$. Note that the peak in (a) is located at $\vect ka=(0.00,1.78)\approx\left(0,\frac{\pi}{\sqrt{3}}\right)$ while in (b) at $\vect ka=(2.09,0.89)\approx\vect q_0a=\left(+\frac{2\pi}{3},\frac{\pi}{2\sqrt{3}}\right)$.
}
\label{numer}
\end{figure}
%%%%%%%%%%%%%%%%%%%%%

\section{Fermions in a triangular lattice}

The frustrated classical magnetism in a triangular optical lattice has been demonstrated experimentally in a Bose system \cite{sengstock11}. Within the mean field approximation Bose-Einstein condensate wavefunction is a Bloch wave with the wavevector corresponding to the minimum of the dispersion relation (\ref{ek}). For the antiferromagnetic interactions the system experiences the frustration, because the tendency of the wavefunction to change the phase by $\pi$, when we jump between neighboring sites, can not be reconciled with the triangular lattice geometry. 

Consider now a mixture of fermions in the different internal states (say spin up $\uparrow$ and down $\downarrow$ states) with the attractive contact interactions in a 2D triangular optical lattice. We assume that the double harmonic modulations of the lattice allows us to adjust any phase of the complex tunneling amplitudes  (\ref{rates}). In the tight-binding approximation the Hamiltonian of the Fermi system reads
\bea
\hat H_F&=&-\sum_{\la ij\ra}J_{ij}\left(\hat a_{i\uparrow}^\dagger\hat a_{j\uparrow}+\hat a_{i\downarrow}^\dagger\hat a_{j\downarrow}\right)-\mu\sum_i\left(\hat n_{i\uparrow}+\hat n_{i\downarrow}\right)
\cr &&
-U\sum_i\hat a_{i\downarrow}^\dagger\hat a_{i\uparrow}^\dagger\hat a_{i\uparrow}\hat a_{i\downarrow},
\eea
where  the operator $\hat a_{i\uparrow}$ annihilates spin-up  fermion at $i$-site, $\hat n_{i,\uparrow}=\hat a_{i\uparrow}^\dagger\hat a_{i\uparrow}$ and similarly for spin-down fermions. The tunneling amplitude $J_{ij}=J^*_{ji}$ and it is equal $J_\alpha$ or $J_\beta$, Eqs.~(\ref{rates}), depending on a direction of the tunneling in the triangular lattice, see Fig.~\ref{trlatt}. The parameter $U>0$ characterizes the inter-species, on-site, attractive interactions and $\mu$ stands for the chemical potential of the Fermi system. 

The standard BCS approach \cite{pethick} leads to the effective Hamiltonian
\bea
H_{F,{\rm eff}}&=&-\sum_{\la ij\ra}J_{ij}\left(\hat a_{i\uparrow}^\dagger\hat a_{j\uparrow}+\hat a_{i\downarrow}^\dagger\hat a_{j\downarrow}\right)-\mu\sum_i\left(\hat n_{i\uparrow}+\hat n_{i\downarrow}\right)
\cr &&
+\sum_i\left(\Delta_i\;\hat a_{i\uparrow}^\dagger\hat a_{i\downarrow}^\dagger+
\Delta_i^*\;\hat a_{i\downarrow}\hat a_{i\uparrow}\right),
\label{efff}
\eea
where the pairing function
\be
\Delta_i=U\la\hat a_{i,\uparrow}\hat a_{i,\downarrow}\ra.
%=U\sum_{\vect k}u_{\vect k}(i)v_{\vect k}^*(i),
\ee
If the phases of the tunneling amplitudes (\ref{rates}) are zero the ground state of the system corresponds to constant pairing function $\Delta_i={\rm const}$.
However, the pairing function can acquire a non-trivial phase when the tunneling amplitudes become complex. In order to find the ground state of the system let us look for the solutions of the Bogoliubov-de~Gennes equations in the form
\be
\left[ \!
\begin{array}{c}
{u}_{\vect k}(\vect r_i) \\ v_{\vect k}(\vect r_i)
\end{array}
\! \right]=
\frac{e^{i\vect k\cdot \vect r_i}}{\sqrt{N_s}}\left[ \!
\begin{array}{c}
U_{\vect k}\;e^{i\vect k_0\cdot \vect r_i}\\ V_{\vect k}\;e^{-i\vect k_0\cdot \vect r_i}
\end{array}
\! \right],
\label{bdgsolut}
\ee
where $U_{\vect k}$ and $V_{\vect k}$ satisfy
\begin{equation}
\left[\!
\begin{array}{cc}
E(\vect k+\vect k_0)-\mu & \bar\Delta \\
\bar\Delta^*  & -\tilde E(\vect k-\vect k_0)+\mu
\end{array}
\! \right]
\!
\left[ \!
\begin{array}{c}
U_{\vect k} \\ V_{\vect k}
\end{array}
\! \right]
=
\varepsilon_{\vect k} \!
\left[ \!
\begin{array}{c}
U_{\vect k} \\ V_{\vect k}
\end{array}
\! \right],
\label{bdgk}
\end{equation}
and $|U_{\vect k}|^2+|V_{\vect k}|^2=1$. In Eqs.~(\ref{bdgk}), $E(\vect k)$ is the dispersion relation (\ref{ek}) while $\tilde E(\vect k)=E(\vect k; \varphi_\alpha\rightarrow -\varphi_\alpha,\varphi_\beta\rightarrow -\varphi_\beta)$.  Solving (\ref{bdgk}) we obtain the following eigenvalues \bea
\varepsilon_{\vect k,\pm}&=&\frac{E(\vect k+\vect k_0)-\tilde E(\vect k-\vect k_0)}{2} \pm {\delta\varepsilon}_{\vect k},
\label{excit1}
\eea
where
\bea
{\delta\varepsilon}_{\vect k}&=&\sqrt{\frac{(E(\vect k+\vect k_0)+\tilde E(\vect k-\vect k_0)-2\mu)^2}{4}+|\bar\Delta|^2}. \cr &&
\eea
The excitation spectrum of the system, i.e. the upper branch $\varepsilon_{\vect k,+}$, may become negative for some $\vect k$. In such a case, the corresponding quasi-particles are present even at zero temperature. Therefore, at $T=0$, the proper equation for $\bar\Delta$ reads 
\bea
\bar\Delta&=&\frac{U}{N_s}\sum_{\vect k}
%U_{\vect k,+}V_{\vect k,+}
\frac{\bar\Delta}{2\;{\delta\varepsilon}_{\vect k}}
[1-2\theta(-\varepsilon_{\vect k,+})],
\eea
where the Heaviside function $\theta(-\varepsilon_{\vect k,+})$ ensures that quasi-particles corresponding to negative energy spectrum are also included \cite{pethick}. Finally the desired pairing function becomes
\bea
\Delta_i&=&U\sum_{\vect k}u_{\vect k,+}(\vect r_i)v_{\vect k,+}^*(\vect r_i)[1-2\theta(-\varepsilon_{\vect k,+})]
\cr &=&e^{i2\vect k_0\cdot \vect r_i}\bar\Delta.
\label{di2k}
\eea

When we switch from $\varphi_\alpha=\varphi_\beta=0$ to $\varphi_\alpha=0$ and $\varphi_\beta\ne 0$ the minimum of the dispersion relation (\ref{ek}) is shifted from $\vect k=0$ to $\vect k=\vect k_0=\left(0,\frac{2\varphi_\beta}{a\sqrt{3}}\right)$. In the ground  state of the system fermions occupy energy levels starting from the new minimum up to the Fermi level. Thus, all fermions acquire quasi-momentum $\vect k_0$ and consequently the pairing function gets the quasi-momentum $2\vect k_0$, see (\ref{di2k}). 

For $|J_\alpha|=|J_\beta|$ and $\varphi_\alpha=\pi$ there are two non-equivalent, degenerate minima of $E(\vect k)$. For example, for $\varphi_\beta=\pi/4$ they are located at $\vect k=\left(\pm\frac{2\pi}{3a},\frac{\pi}{2a\sqrt{3}}\right)$, see Fig.~\ref{disrel}c. In the ground state, fermions occupy energy levels with quasi-momenta around both of the minima. {  
The non-zero pairing function exists for different values of $\vect k_0$. 
However, for $\vect k_0=\left(0,\frac{\pi}{2a\sqrt{3}}\right)$ we obtain the lowest energy of the Hamiltonian (\ref{efff}). 
A slight change of $\vect k_0$ causes a rapid decrease of the energy gap in the excitation spectrum (\ref{excit1}). } 
In Fig.~\ref{numer}a we present Fourier transform $|\Delta_{\vect k}|^2$ of the pairing function, where $\Delta_{\vect k}=\sum_{i}\Delta_ie^{-i\vect k\cdot\vect r_i}/\sqrt{N_s}$, obtained numerically for a finite system. For the chemical potential $\mu=0$, that corresponds to half filling regime for non-interacting particles, $|J_\beta|/|J_\alpha|=1$, $U/|J_\alpha|=2$ we obtain $|\Delta_i|/|J_\alpha|=0.111 $ at the center of the lattice which agrees with the analytical solution $\bar\Delta/|J_\alpha|=0.109$ for infinite lattice. In Fig.~\ref{numer}a we see that even for the lattice of $60\times 60$ sites there exists a clearly resolved peak at $\vect k\approx2\vect k_0=\left(0,\frac{\pi}{a\sqrt{3}}\right)$. 

{ 
It was discussed \cite{sengstock11}, Bose systems can simulate the frustrated classical magnetism. We show that the similar phenomenon can be simulated in the Fermi system. Indeed, in triangular lattice with complex tunnelings, the phase of the complex pairing function is the one that is mapped onto orientation of classical spins. } 

%In the Fermi system the phase of the pairing function can be mapped onto orientations of classical spins. 
%We see that, in triangular lattice with complex tunnelings, the Cooper pairs can simulate the frustrated magnetic systems as well, when the phase of the pairing function of the Fermi system is mapped onto orientations of classical spins. 

\section{Bose-Fermi mixture in a triangular lattice}

In this section we consider a situation when the fermions coexist with  molecular dimers -- pairs of spin-up and spin-down fermions. The dimers form a Bose-Einstein condensate. Such a mixture can be prepared by sweeping the system over a Feshbach resonance that creates a molecular BEC and leaves some fraction of unbound, repulsively interacting fermions. Then crossing a second Feshbach resonance one is able to change the interactions between fermions from repulsive to attractive, turn  unbound fermions into BCS pairs \cite{niederberger09}. The process does not affect the molecules at the same time. For this purpose Feshbach resonances at 202~G and 224~G for $^{40}$K atoms \cite{regal03} seem to be quite suitable. We also assume the presence of a weak coupling that transforms dimers into unbound fermions and vice versa. It can be realized via the photo-dissociation and photo-association. For a large molecular BEC the weak coupling does not influence significantly the condensate wavefunction and therefore we neglect dynamics of the BEC. The system under our consideration can be reduced to the following Hamiltonian
\be
\hat H=\hat H_F+\hat H_{BF},
\ee
with
\be
\hat H_{BF}=\gamma\sum_i \left(\psi_i^*\;\hat a_{i\downarrow}\hat a_{i\uparrow}+\psi_i\;\hat a_{i\uparrow}^\dagger\hat a_{i\downarrow}^\dagger\right),
\label{hhbf}
\ee
where the BEC wavefunction $\psi_i=\sqrt{n_B}e^{i\vect q_0\cdot \vect r_i}$ is the ground state solution for bosons in the triangular lattice, i.e. $\vect q_0$ corresponds to the minimum of the dispersion relation (\ref{ek}).
{  For simplicity reasons but without loss of generality we choose the same dispersion relation for molecules and for fermions. In the system under consideration, the tunneling amplitudes for molecules in a shaken optical lattice depend on a molecular state populated in the photo-association process. The details of this process are not considered in the present paper.} 
The coupling constant $\gamma$ characterizes transfer  of dimers into unbound fermions and vice versa. We consider real $\gamma\ge 0$.

In the presence of the condensate of dimers the BCS effective Hamiltonian (\ref{efff}) has to be supplemented with (\ref{hhbf}), that is
\be
\hat H_{\rm eff}=\hat H_{F,{\rm eff}}+\hat H_{BF}.
\ee
{  

In the presence of bosons, if $\vect k_0=\vect q_0/2$, the simple analytic solution (\ref{bdgsolut}) of the corresponding 
Bogoliubov-de~Gennes equations exists. This solution need not correspond to the ground 
state of the system. However, we will see that for sufficiently strong coupling 
between bosons and fermions such a solution becomes the ground state solution.}
Employing (\ref{bdgsolut}) with $\vect k_0=\vect q_0/2$ we obtain the following
equation for $\bar\Delta$
\bea
\bar\Delta&=&\frac{U}{N_s}\sum_{\vect k}
%U_{\vect k,+}V_{\vect k,+}
\frac{\bar\Delta+\gamma\sqrt{n_B}}{2\delta\varepsilon_{\vect k}}
[1-2\theta(-\varepsilon_{\vect k,+})],
\label{gapb}
\eea
where, in the present case, the excitation spectrum is 
\bea
\varepsilon_{\vect k,+}&=&\frac{E(\vect k+\vect q_0/2)-\tilde E(\vect k-\vect q_0/2)}{2}+\delta\varepsilon_{\vect k},
\eea
with
\bea
\delta\varepsilon_{\vect k}&=&\left[\frac{(E(\vect k+\vect q_0/2)+\tilde E(\vect k-\vect q_0/2)-2\mu)^2}{4}\right. 
\cr && \left. +|\bar\Delta+\gamma\sqrt{n_B}|^2\right]^{1/2},
\cr &&
\eea
and the resulting pairing function 
\be
\Delta_i=e^{i\vect q_0\cdot \vect r_i}\bar\Delta.
\label{delq0}
\ee

Let us concentrate on the triangular lattice with $|J_\beta|/|J_\alpha|=1$, $\varphi_\alpha=\pi$ and $\varphi_\beta=\pi/4$ that corresponds to the dispersion relation plotted in Fig.~\ref{disrel}c. The dispersion relation reveals the two non-equivalent minima, but solution of the Gross-Pitaevskii equation for bosons chooses the Bloch wave with the quasi-momentum corresponding to  one of the minima. The signatures of such a spontaneous symmetry breaking are observed experimentally \cite{sengstock11}. We assume that Bose system chooses $\vect q_0=\left(+\frac{2\pi}{3a},\frac{\pi}{2a\sqrt{3}}\right)$ and  analyze its influence on the Fermi system. 

 We consider the system with $\mu=0$.
If $\gamma=0$ the Cooper pairs with the quasi-momentum $\vect q_0$ do not exist, i.e. $\bar\Delta=0$ is the only solution of (\ref{gapb}). If the coupling between bosons and fermions is present, but $\gamma\sqrt{n_B}/|J_\alpha|< 2.112$, the system reveals gap-less superfluidity \cite{pethick}. The Cooper pairs with the quasi-momentum $\vect q_0$ appear ($\bar\Delta\ne 0$), but there is no energy gap in the  excitation spectrum.  The system possesses  quasi-momenta $\vect k$ for which the excitation energies $\varepsilon_{\vect k,+}<0$ and consequently the corresponding quasi-particles are present even at $T=0$. {  Concerning the ground state of the system, numerical solutions of the Bogoliubov-de~Gennes equations are analyzed. It is found that increasing parameter $\gamma$ causes a gradual enlargement of the peak at $\vect k=\vect q_0$ in the Fourier transform of the pairing function together with a reduction of the peak at $\vect k=\left(0,\frac{\pi}{a\sqrt{3}}\right)$ (the solution in the absence of bosons considered in the previous section). For $\gamma\sqrt{n_B}/|J_\alpha|\approx 0.3$ we observe a crossover, the peak at $\vect k=\left(0,\frac{\pi}{a\sqrt{3}}\right)$ becomes hardly visible and the ground state starts to be well reproduced by the paring function (\ref{delq0}).}

For $\gamma\sqrt{n_B}/|J_\alpha|\ge 2.112$ the energy gap  shows up, $\varepsilon_{\vect k,+}>0$. There is no quasi-particle at zero-temperature and the pairing function (\ref{delq0}) is related to the ground state of the system.
In Fig.~\ref{numer}b we show the Fourier transform of the pairing function obtained numerically for the triangular lattice of $60\times 60$ site where the strong peak at $\vect k\approx\vect q_0$ is clearly visible. The pairing function at the center of the lattice is $|\Delta_i|/|J_\alpha|=0.891$  and the energy gap in the excitation spectrum is $0.187|J_\alpha|$. Those numbers agree with the solutions for infinite system, i.e. $\bar\Delta/|J_\alpha|=0.891$ and min$(\varepsilon_{\vect k,+})=0.194|J_\alpha|$.

Thus, we can describe the behaviour of the system in the following way. In a triangular optical lattice with complex tunnelings we are able to realize a BEC in the ground state with the wavevector located at an arbitrary position in the reciprocal space. If the superfluid fermions are also present in the lattice and there is the sufficiently strong coupling between fermions and bosons, the phase of the BCS pairing function reflects the phase of the BEC wavefunction. 

%We have concentrated on a zero temperature case. However, even for $T>0$ the BCS approach shows that the pairing function in the configuration space still follows the phase of the condensate wavefunction, Eq.~(\ref{deltai}). Indeed, thermal effects reduce $\Delta_{\vect k}$ but do not influence the relation (\ref{deltai}). Thus, as long as the superfluidity exists, the BCS pairing function can mimic the frustrated classical magnetism.

\section{Conclusions}

In summary, we have shown that the time-reversal symmetry breaking in an optical lattice potential allows us to realize complex tunneling amplitudes 
in the corresponding tight-binding model. We have considered a simple scheme of the symmetry breaking by means of two harmonic 
modulations of the lattice, but the generalization to more complicated modulations is straightforward. 

We have studied a fermionic system as well as a Bose-Fermi mixture in a triangular lattice potential with complex tunnelings. 
In such a lattice the Bose system can simulate the frustrated classical magnetism \cite{sengstock11}. We have shown that this behaviour is similar for fermions where the pairing function acquires a complex phase.
Assuming the presence of a coupling mechanism -- an exchange of unbound fermions and bosonic molecules,  
we have shown that the complex phase of Bose wavefunction is mapped to fermions as reflected  in the Fermi pairing function.

While preparing the second, decompressed version of this manuscript we became aware of a very recent preprint \cite{struck12} where the authors consider similar idea for realization of complex tunneling amplitudes in Bose systems both theoretically and experimentally. 

\section{Acknowledgements}

The work of K.S. and  K.T. was financed from Polish National Center for Science funds received through decisions DEC-2011/01/B/ST3/00512 and DEC-2011/01/N/ST2/00424,  respectively. J.Z. acknowledges support within Polish Government scientific funds for 2009-2012 (research project).

\end{document}